# Positive Energy Wormholes Made from Normal Charges

Ron Lenk*

May 12, 2010

Abstract

In this article we look at the class of positive-energy density wormholes introduced by Eiroa and Simeone, and ask the question "How close can we come to building this wormhole with normal, electrically charged matter?" Using a series expansion, we determine that the stress-energy tensor of the metric can be made arbitrarily close to that of a charged line-source by taking the exponent of the radial coordinate to be sufficiently small. We also find terms in the expansion which resemble QED radiative corrections to the Coulomb field of an electron.

As is well-known, wormholes necessarily must have negative energy associated with them [1]. However, recently a new class of wormholes has been introduced by Eiroa and Simeone [2]. They avoid the usual exotic matter requirements for wormholes by requiring only that the geodesics restricted to a plane normal to the symmetry axis flare out, as proposed in [3], rather than that all geodesics flare out. This allows them to have everywhere positive energy density for this type of wormhole, and they are thus potentially constructible. The goal of this paper is to establish how close these metrics can be to ones constructed from ordinary charged matter.

The general form of the metric considered is given by [4]

$$ds^2 = \rho^{2m^2} H^2 \left(d\rho^2 + dz^2\right) + \rho^2 H^2 d\varphi^2 - H^{-2} dt^2 \tag{1}$$

with

$$H = k_1 \rho^m + k_2 \rho^{-m} \tag{2}$$

* e-mail: ron.lenk@reliabulb.com

with $k_1$, $k_2$ and m real, but not necessarily positive, constants. The units of $k_1$ are length$^{1/m}$ and of $k_2$ are length$^m$, so that H is dimensionless. To construct the wormhole, the volume inside a ball of radius a is excised from both of two copies of this metric, and the two boundaries at ρ = a are glued together to form a wormhole with throat radius a.

This will represent a positive energy wormhole if the two conditions

$$2a\left(k_1a^m + k_2a^{-m}\right)\left[(1+m)k_1a^m + (1-m)k_2a^{-m}\right] > 0 \tag{3a}$$

$$2a^{2m^2+1}\left(k_1a^m + k_2a^{-m}\right)^3\left[(1+m)^2k_1a^m + (1-m)^2k_2a^{-m}\right] < 0 \tag{3b}$$

are met. As shown in Ref. [2], a necessary (but not sufficient) condition for the second inequality to be satisfied is for $|m| < 1$, which we will assume from now on.

The stress-energy tensor for Eq. (1) is

$$T^{\mu}_{\nu} = -8\pi diag(Q,Q,-Q,-Q)$$

$$Q = \frac{k_1k_2m^2}{2\pi H^4\rho^{2m^2+2}} \tag{4}$$

in units with G = c = 1. It is trace-free, as appropriate for a massless field. We want this to approximate as closely as possible the stress-energy tensor for a charged line-source. Since for a charged line E = 2λ / ρ, with λ the line charge density, the stress-energy tensor is as in Eq. (4), with

$$Q_{CL} = -\frac{\lambda^2}{2\pi\rho^2} \tag{5}$$

Expanding the denominator of Q in Eq. (4), we find five terms in differing powers of ρ, with numerical factors that sum to 16. We select

$$k_1k_2m^2 = -16\lambda^2 \tag{6}$$

where the extra factor of 16 is to cancel the $H^4$ terms, and we want

$$k_1^{\,4}\rho^{2(m+1)^2} + 4k_1^{\,3}k_2\rho^{2m^2+2m+2} + 6k_1^{\,2}k_2^{\,2}\rho^{2m^2+2} + 4k_1k_2^{\,3}\rho^{2m^2-2m+2} + k_2^{\,4}\rho^{2(m-1)^2} \tag{7}$$

to be as close to $\rho^2$ as possible. If we try to make one of the exponents in Eq. (7) equal to 2, direct solution of the equations show that the choices for m are 0, ±1, and ±2. Since we must have $|m| < 1$ in order to have a positive surface energy density on the shell, this would only leave m = 0, which is a flat space-time.

The next best choice is to look at small deviations from $1 / \rho^2$. We try m = $\pm\varepsilon$, $1 - \varepsilon$ and $-1 + \varepsilon$, with $\varepsilon$ small. The choices with m = $1 - \varepsilon$ and $-1 + \varepsilon$ for Q give powers of $\rho$ to the fourth, sixth and eighth, in addition to $\rho^2$. Since our goal is to closely match Eq. (5), we end up selecting m = $\pm\varepsilon$, which makes all of the terms in the denominator of Q close to $\rho^2$. For example, with m = 0.01 we have

$$Q = -\frac{8\lambda^2}{\pi}\frac{1}{\rho^{2.0402} + 4\rho^{2.0202} + 6\rho^{2.0002} + 4\rho^{1.9802} + \rho^{1.9602}} \tag{8}$$

As shown in Fig. 1, for a = 0.001, the difference between this expression and the ideal charged line-source is less than 1% out to $\rho = 1$.

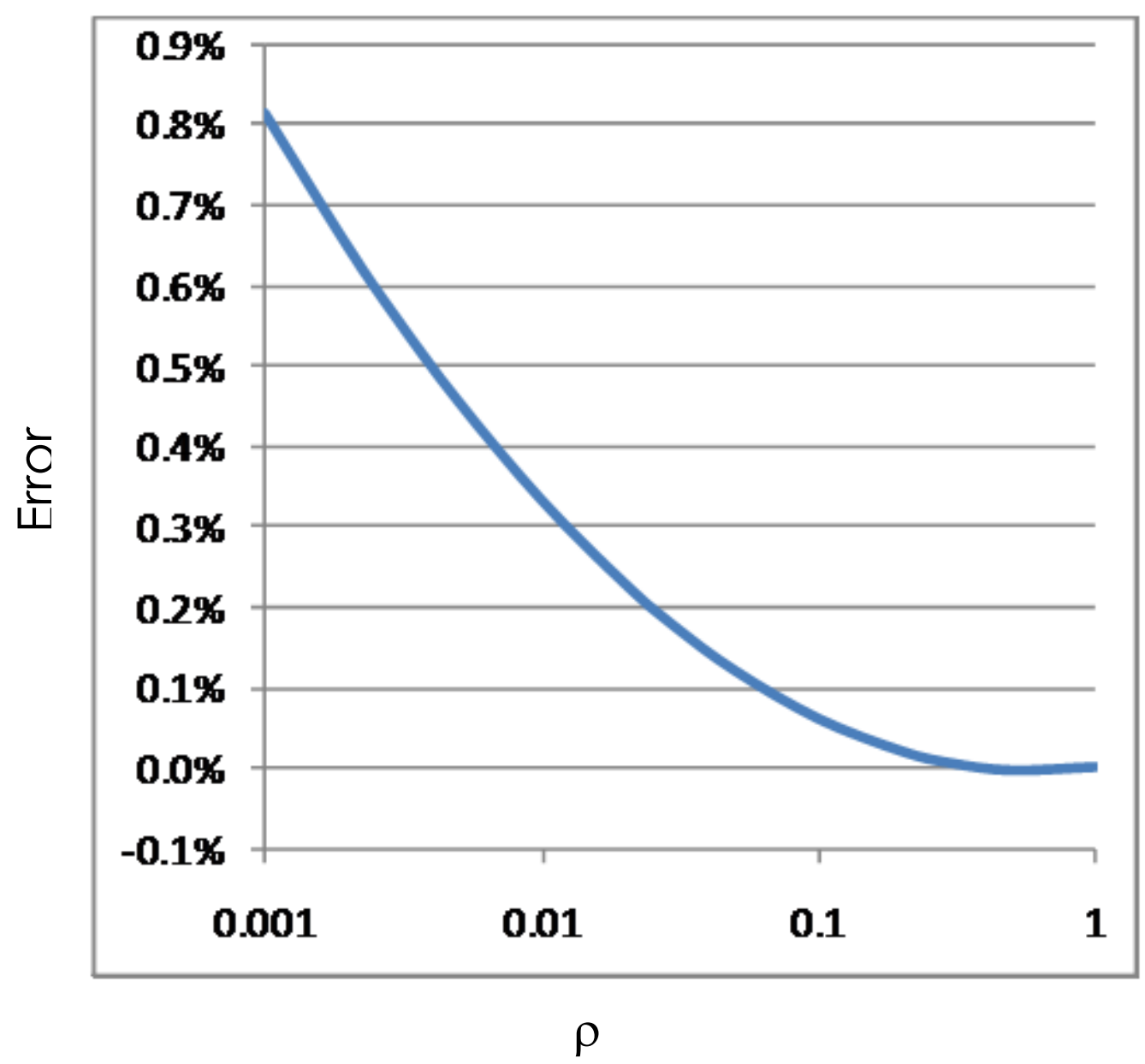

Figure 1: The difference between $Q_{CL}$ and Q as a function of $\rho$ for Eq. (8). For $\rho > 0.001$, the error is less than 1%.

The series expansion yields

$$Q = \frac{k_1 k_2 m^2}{2\pi}\frac{1}{\rho^2 (k_1 + k_2)^4}\left[1 + \frac{4(k_1 - k_2)\ln(\rho)}{(k_1 + k_2)} m + O(m^2)\right] \tag{9}$$

If we now take

$$\frac{k_1 k_2 m^2}{(k_1+k_2)^4} = -\lambda^2 \tag{10}$$

then to lowest order in m this matches the charged line-source of Eq. (5). We can also take a power series in m for Eqs. (3a) and (3b),

$$2a(k_1+k_2)^2 + 2a(k_1+k_2)(k_1-k_2)[2\ln(a)+1]m > 0 \tag{11a}$$

$$2a(k_1+k_2)^4 + 4a(k_1+k_2)^3(k_1-k_2)[2\ln(a)+1]m > 0 \tag{11b}$$

showing that in the limit $m \to 0$ these conditions can be satisfied for any values of $k_1$ and $k_2$.

Finally, it is interesting to note that the second term of the expansion for the stress-energy tensor, Eq. (9), has a ρ-dependence of $\ln(\rho) / \rho^2$. This term is interestingly reminiscent of the radiative correction term to the Coulomb field of an electron in QED [5]. The third order term turns out to have a $\ln^2(\rho) / \rho^2$ dependence, which is similar to the 'double-logarithmic' radiative correction term in QED [5]. Work is in progress to see if a QED model of a charged spherical source might include similar terms as well.

## Acknowledgment

The author is grateful to D. Singleton for assistance.